\documentclass[journal]{IEEEtran}
\usepackage[T1]{fontenc}
\usepackage{graphicx}
\usepackage{pifont} 
\usepackage[normalem]{ulem} 
\usepackage{amssymb}
\usepackage{indentfirst} 
\usepackage{setspace}
\usepackage{amsmath} 
\usepackage{float}
\usepackage[oldenum,olditem]{paralist}

\usepackage[hyphens]{url}
\usepackage[dvipsnames]{xcolor}
\usepackage{tabularx}

\newcolumntype{x}[1]{%
>{\centering\hspace{0pt}p{#1}}}%

\begin{document}
\title{On the Privacy of National Contact\\Tracing COVID-19 Applications:\\The \textit{Coronavírus-SUS} Case}

\author{
	\IEEEauthorblockN{
	J\'eferson Campos Nobre,
    Laura Rodrigues Soares,
	Briggette Olenka Roman Huaytalla,\\
    Elvandi da Silva J\'unior,
	Lisandro Zambenedetti Granville\\
	}
	\IEEEauthorblockA{
		Federal University of Rio Grande do Sul, Brazil\\
		Email: \{jcnobre, lrsoares, borhuaytalla, elvandi.junior, granville\}@inf.ufrgs.br 
	}
}

\maketitle



\begin{abstract}

The 2019 Coronavirus disease (COVID-19) pandemic, caused by a quick dissemination of the Severe Acute Respiratory Syndrome Coronavirus 2 (SARS-CoV-2), has had a deep impact worldwide, both in terms of the loss of human life and the economic and social disruption. The use of digital technologies has been seen as an important effort to combat the pandemic and one of such technologies is contact tracing applications. These applications were successfully employed to face other infectious diseases, thus they have been used during the current pandemic. However, the use of contact tracing poses several privacy concerns since it is necessary to store and process data which can lead to the user/device identification as well as location and behavior tracking. These concerns are even more relevant when considering nationwide implementations since they can lead to mass surveillance by authoritarian governments. Despite the restrictions imposed by data protection laws from several countries, there are still doubts on the preservation of the privacy of the users. In this article, we analyze the privacy features in national contact tracing COVID-19 applications considering their intrinsic characteristics. As a case study, we discuss in more depth the Brazilian COVID-19 application \textit{Coronavírus-SUS}, since Brazil is one of the most impacted countries by the current pandemic. Finally, as we believe contact tracing will continue to be employed as part of the strategy for the current and  potential future pandemics, we present key research challenges.

\end{abstract}

\begin{IEEEkeywords} 
Contact Tracing; Coronavírus-SUS; COVID-19; SARS-CoV-2; Privacy.

\end{IEEEkeywords}

\IEEEpeerreviewmaketitle

\section{Introduction}


The current global health crisis -- \textit{i.e.}, the 2019 Coronavirus disease (COVID-19) pandemic -- was caused by the spread of the Severe Acute Respiratory Syndrome Coronavirus 2 (SARS-CoV-2) \cite{guidelines-who}. This virus was discovered in late 2019 in China, from where it quickly disseminated worldwide. The form of transmission is through contaminated secretions that spread through the air or contact with a contaminated object. This transmission can occur even before the manifestation of the first symptoms. Currently, most countries are unable to carry out tests on their entire population in a short time because it requires large financial investment, qualified personnel, and the manufacture of products. In this context, governments and the private sector come together to attempt to take advantage of digital technologies in efforts to combat the pandemic \cite{ferretti2020quantifying}.


Contact tracing\footnote{In this article, we use ``contact tracing" as a synonym for ``digital contact tracing"} is a technological method of monitoring the progress of an infectious disease on a large population by detecting the contacts between infected and healthy individuals \cite{EC}. These contacts are traced through applications (apps) installed on mobile devices, usually smartphones. These apps use different techniques to trace proximity: Global Positioning System (GPS), triangulation of cellular operator antennas, electronic transaction data (\textit{e.g.}, credit card data), and Bluetooth. In addition, tracing data (\textit{i.e.}, contact history) can be treated in a centralized manner (processed and stored on a central entity) or in a decentralized one (using users' devices for processing and storage).


Contact tracing apps have been employed to face the COVID-19 pandemic since they were proven to be effective in previous pandemics \cite{Danquah2019}. Also, the huge adoption of smartphones along with the evolution of mobile network coverage has increased the number of users who can use such apps. In addition, the Google/Apple Exposure Notification (GAEN) system \cite{gaen-google} fostered the development of contact tracing apps by health authorities in a well-defined manner, for both Android and iOS devices. Thus, contact tracing rose to fame during the COVID-19 pandemic \cite{ferretti2020quantifying}, being implemented by some countries at different moments of the pandemic and with varying penetration rates.


The use of contact tracing must be accompanied by a discussion on how to preserve the privacy of users. Otherwise, users may not adopt contact tracing apps if they are suspicious about the private information that is being shared or, depending on the users' privacy concerns, even leaked. As such, contact tracing apps should be designed to avoid personally identifiable information to be collected on user/device basis. Data protection laws from several countries impose restrictions on the approach to perform data processing along with the details on the security measures required on stored personal data. On the other hand, health agencies are usually the only ones who can operate on collected data. In any case, there are significant concerns regarding an unrestricted digital monitoring (\textit{i.e.}, mass surveillance) by authoritarian governments, which can ultimately lead to ``orwellian" systems. 


In this article, we analyze privacy features in national contact tracing apps designed to combat the COVID-19 pandemic. Because of the emergence of such apps and the impact of similar ones on other pandemics, it is possible to conclude that contact tracing will continue to be employed as part of the SARS-CoV-2 containment strategies. However, logging encounters and exposure notifications may lead to undesirable privacy attacks, especially considering large-scale deployments. We depict a perspective on the privacy of national contact tracing COVID-19 apps considering several variables, such as the number of users, penetration, and localization. Finally, given that Brazil is one of the most impacted countries by the current pandemic, we evaluate in more detail the Brazilian COVID-19 application \textit{Coronavírus-SUS}.


This article is organized as follows. In the next section, we present a background on contact tracing, to then provide, in the following section, a general view of privacy in the context of COVID-19. Afterwards, several contact tracing apps employed by national authorities around the world are evaluated. We then present a case study regarding the \textit{Coronavírus-SUS} app. Finally, we conclude the article in presenting key research challenges and final remarks.

\section{Contact Tracing in a Nutshell}
\label{sec:background}


National governments as well as the private sector are working towards developing computational tools to help on the effective management of the COVID-19 pandemic. Contact tracing apps help break the chain of virus transmission by notifying users whenever they are close to infected individuals, thus before they become prone to infect others \cite{anglemyer2020digital}. There are precedents for such apps being used on other health crises, such as Ebola, Tuberculosis, and HIV, as a part of the strategy for disease outbreak control \cite{Danquah2019}. Consequently, even in the early stages of the COVID-19 pandemic, it was hypothesized that contact tracing apps could be an important tool to decrease the basic reproduction number ($R_{0}$) of SARS-CoV-2.

\begin{figure}[ht]
  \centering
  \includegraphics[width=.45\textwidth]{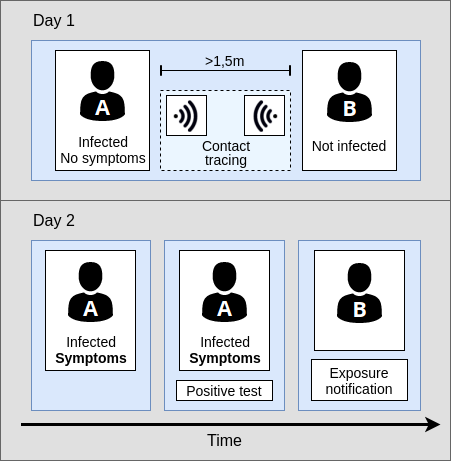}
  \caption{Basic operation of contact tracing applications.}
  \label{fig:basic}
\end{figure}

Contact tracing apps enable the monitoring of interactions of infected and healthy users, thus detecting potential infections. After such detection, it is necessary to notify the users, so measures to limit the spread can be taken either by the user himself or by the  health authorities. The tasks performed by these apps can be aggregated into 3 groups: detection of contact events (proximity tests), transmission, and exposure notification. The basic operation of contact tracing apps is depicted in Figure \ref{fig:basic}. In Day 1, user A (infected and with no symptoms) has an contact with user B (healthy). In Day 2, after symptoms onset on user A, he is tested and shares the positive test result with the app which triggers an exposure notification to user B.

The detection of contact events can be performed through several methods and technologies. The majority of these methods is related with sensor technologies that are integrated into current mobile devices. Proximity tracing is a method usually performed using Bluetooth Low Energy (BLE) to transmit messages containing identifiers to nearby devices. This can be enhanced with the use of ultrasound so as to achieve better accuracy. Location tracking is a method that can be performed using data from the Global Positioning System (GPS) or cell tower triangulation. Geotagging is a method that can also be used, where users scan the QR code with their mobile device to record their visits, and consequently their localization data. Other additional methods include the use of consumer credit card data and Closed-Circuit TeleVision (CCTV) with facial recognition.

\begin{figure*}[ht]
  \centering
  \includegraphics[width=\textwidth]{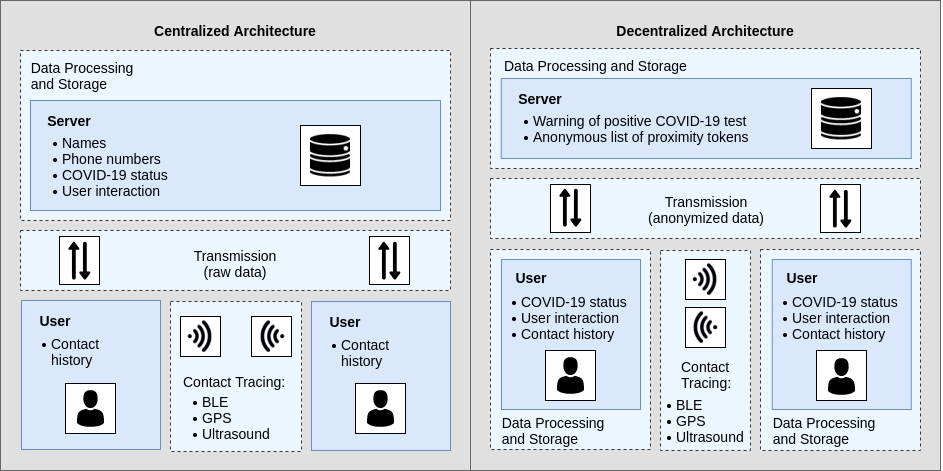}
  \caption{Centralized and decentralized architecture for contact-tracing applications.}
  \label{fig:architecture}
\end{figure*}

 
Different architectures can be employed to collect users' data and contact events. These architectures are depicted on Figure \ref{fig:architecture}. Centralized architectures collect the raw contact history data of mobile phones through a communication network. After that, this data is stored and processed in a central server, which generates reports and sends exposure notifications using the same network. However, this centralization brings concerns over dependability and performance. Decentralized architectures, as can be seen in Figure \ref{fig:architecture}, employ local resources for data storage and processing, which is feasible since only contact events of the last 14 or 21 days (the number of days depends on the app) need to be preserved. In both architectures, mobile devices are usually responsible for generating temporary (also called ephemeral) identifications.  

Google and Apple formed a partnership to develop an interoperable interface (between Android and iOS) contact event detection based on BLE technology in April 2020. The result of this partnership is the Google/Apple Exposure Notification (GAEN) system, which presents an Application Programming Interface (API). This system is implemented at the operating system level to avoid privilege problems. Contact tracing apps from many countries use this API for their notification of exposure and the generation of temporary tokens (to preserve user privacy). However, GAEN API is not open source and its public documentation is limited, which brings concerns about the transparency of the API itself \cite{leith2020gaen}.


A contact tracing app collects and exchanges sensitive users' data on a regular basis. However, such ability comes at a cost: privacy concerns. For example, the use of proximity tracing causes the smartphone to send messages in a repetitive manner that can be inspected. Besides, characteristics of different architectures can impact the preservation of users' privacy, since the transmission, processing, and storage of users' data are performed in distinct ways. For instance, the anonymization and encryption can be performed either in the mobile device or in the server.

\section{Privacy in the Context of COVID-19}
\label{sec:privacy}


The COVID-19 pandemic spread worldwide quickly and aggressively and exposed either the lack or precariousness of public health policies to combat epidemics. Most countries are unable to carry out tests on the entire population at first notice, as they require a large financial investment, qualified personnel, and the manufacture of products. Before large-scale vaccination, social distancing and isolation still remain as the most effective measures against the spread of COVID-19 \cite{guidelines-who}. In addition to the health of the population, this pandemic also brought new risks to the security of personal data, as the world adapts to the novel paradigms for working and socializing. Threats to the privacy of personal data are one of the main issues and are sure to be felt for years to come if not addressed appropriately from the beginning.


The privacy subject comes under scrutiny almost immediately as the majority of the global workforce transitioned to home office due to social distancing. Contact tracing technologies represent a new incursion into privacy issues. In healthcare applications, the data traveling through potentially insecure connections are especially sensitive, since this concerns the physical welfare of people. Therefore, it is important that technology companies, healthcare providers, and public health officials operate with the highest ethical standards, in particular when the existing privacy regulations do not apply. 


COVID-19 contact tracing apps have another relevant issue besides the sensitive nature of its data: the extensive coverage among the citizens of a country to reach the intended efficacy. Thus, most of the such apps so far are developed by governmental offices \cite{MIT-health-house}. 
This arrangement has several benefits: in most cases, the government already has in place the necessary structure to launch a nation-wide application, as well as the data protection mechanisms to ensure the privacy of its users. However, even institutional offices in a country are not immune to data leakage. For example, in November 2020, an employee error disclosed passwords from the Ministry of Health of the Brazilian Government, leading to massive data leakage regarding COVID-19 patients \cite{estadao}. 
At least 16 million people have had their data exposed, including pre-existing illnesses and COVID-19 status. Also, the passwords still gave access to two federal government records: one for notifications of suspected cases and the other with hospitalizations for Severe Acute Respiratory Syndrome (SARS).


Several data protection regulations were created to ensure the safe handling of sensitive personal information. The Health Insurance Portability and Accountability Act (HIPAA) \cite{hipaa1996} from the United States, a pioneer in data regulation, was created in 1996 to regulate how personally identifiable information from healthcare industries should be maintained. More recently, broader regulation laws were also developed. From the European Union (EU), the General Data Protection Regulation (GDPR) \cite{gdpr2016} aims to protect an individual's personal data not only inside the area but in any enterprise that processes the personal information of its citizens as well. The GDPR was later used as a base for the Brazilian \textit{Lei Geral de Proteção de Dados (LGPD)} \cite{lgpd2018}, Brazil's data protection regulation law. 


 
Contact tracing apps can be a corner case for privacy in the context of COVID-19. These apps handle personally identifiable data (\textit{e.g.}, names, phone number, address, COVID-19 tests), exposure notifications, and graphs of social density (indications of close contact between users). Since the effectiveness of such apps depends on these data, the discussion on the users privacy tends to be put in opposition of public health. Besides, there are concerns that the current state of epidemiological surveillance can surpass the COVID-19 pandemic and become a permanent part of governmental strategies. In next section, we analyse the current landscape of national contact tracing COVID-19 apps and their privacy features. 

\section{National Contact Tracing COVID-19 Apps and Privacy}
\label{sec:national}


The majority of national governments have employed computational-based approaches to face the COVID-19 pandemic. Regarding the specificities of this pandemic, national strategies are of paramount importance because only they can provide the required level of resources and logistics to avoid the shortage of healthcare facilities (especially Intensive Care Unit beds). In this context, the amount of sensible data collected from the users in these approaches is significant. In addition, as discussed before, the necessary social distancing protocols increased concerns over privacy. Such concerns are even more distressful considering that mobile apps have access to users' location. This is exactly the main feature required for contact tracing apps.


In Table \ref{tab:apps}, we survey a selection of contact tracing apps in different countries (For a more comprehensive view, we suggest the ``MIT Technology Review Covid Tracing Tracker" \cite{MIT-health-house}). As prerequisites, the applications should be already launched and from a governmental developer. If these requisites are met, we aimed for a broader set of features such as user number, adherence, and chosen technologies. \textit{Aarogya Setu} from India is by far the largest in number of users, a relevant matter for the present article since Brazil has the 6th largest population in the world. \textit{Ehteraz} from Qatar is the most adopted contact tracing app, reaching 91\% of the population. Israel's \textit{HaMagen} and UK's \textit{NHS COVID-19 App} have satisfactory adherence. In addition, the National Health Service (NHS), England's publicly funded healthcare system, is very similar to Brazil's \textit{Sistema Único de Saúde}. Both New Zealand and Japan have unique approaches to face the pandemic. Mexico, in turn, faces several of the same socioeconomic problems historically found in Brazil. 

\begin{table*}[h]
\centering
\begin{tabular}{lllllll} 
\hline
Location    & Name    & Users    & Penetration & Contact Detection  & Distribution \\ 
\hline\hline
India    & Aarogya Setu    & 163.000.000    & 12,05\%    & Bluetooth, Location     & Centralized \\ 
\hline
Israel    & HaMagen    & 2.000.000    & 22,51\%    & Location   & Centralized \\ 
\hline
Japan    & COCOA    & 7.700.000    & 6,09\%    & Bluetooth, GAEN &     Decentralized \\ 
\hline
Mexico    & CovidRadar    & 50.000    & 0.04\%    & Bluetooth               & Centralized   \\ 
\hline
New Zealand    & NZ COVID Tracer    & 588.800    & 12,10\%    & Bluetooth, QR codes    & Centralized \\ 
\hline
Qatar    & Ehteraz    & 2.531.620    & 91\%    & Bluetooth, Location     & Centralized    \\ 
\hline
UK    & NHS COVID-19 App    & 19.000.000    & 28,51\%    & Bluetooth, GAEN    & Decentralized \\
\hline
\end{tabular}
\caption{Selected national contact tracing apps.}
\label{tab:apps}
\end{table*}


BLE is the most used technology for the detection of contact events in the analyzed applications (Table \ref{tab:apps}). Geo-location technology is used exclusively in Israel's \textit{HaMagen} app, and in conjunction with BLE in \textit{Aarogya Setu} of India and \textit{Ehteraz} of Qatar. Geo-location and a centralized architecture for the data collection data are considered more invasive, because they expose the personal and location data collected from users to several national control authorities. They are also more susceptible to hacks and leaks of personal data. Data centralization is used in applications in India, Israel, Mexico, New Zealand and Qatar. In contrast, applications from Brazil, Japan and the UK use data decentralization. This method is considered less invasive, as the collected data is stored on users' devices, maintaining anonymity. 


The preservation of users' privacy, especially considering the national coverage of the contact tracing apps depicted on Table \ref{tab:apps}, must follow best practices (as well as data protection laws) for the use of population tracking data. In this context, when using data from a source other than the user's own device (\textit{e.g.}, telecommunications operators), the usage of such apps must be performed on a voluntary basis. Thus, these apps should be only allowed to collect personal information strictly related to SARS-Cov-2 tracking. For example, BLE-enabled apps should be prohibited from collecting location data (which is not the case for \textit{Aarogya Setu} and \textit{Ehteraz}). In addition, consent must be provided by the user from health and safety agencies to share a positive test result.
 

Several of the privacy issues found in these applications can also be found in the Brazilian application \textit{Coronav\'irus-SUS}. We purposefully left \textit{Coronav\'irus-SUS} outside the selection so we can conduct a more detailed  study about particular privacy issues in the next section, as well as an overview of the obstacles for contact tracing adhesion in Brazil.

\section{Brazilian Contact Tracing App: Coronavírus-SUS} 
\label{sec:use_case}



The official application for COVID-19 contact tracing in Brazil, \textit{Coronav\'irus-SUS}, was launched on February 28\textsuperscript{th}, 2020 as a basic version, with features such as information about symptoms, COVID-19-related official news, and a map with the public health facilities in one's region. Later, on the September 31\textsuperscript{st}, the app was upgraded with contact tracing functions. \textit{Coronav\'irus-SUS} was promoted by the Brazilian National Government and developed by DATASUS, the Department of Informatics of Brazil's public-funded healthcare system, the \textit{Sistema \'Unico de Sa\'ude} (SUS). It uses GAEN API for anonymous Bluetooth sharing of tokens.



The users of \textit{Coronav\'irus-SUS} are informed about the application's privacy policy when first downloading it or after upgrading to the contact tracing version. That policy states that no personal data is collected, no GPS data is used, all communications are encrypted, all information is stored in data servers in Brazil, and there is no way of finding out one's identity or the identity of whoever comes in contact with. It is possible to use other features (\textit{e.g.}, news and health facility map) and decline to use the contact tracing feature, but the user must agree with the privacy policy anyway.

\begin{figure}[ht]
  \centering
  \includegraphics[width=.5\textwidth]{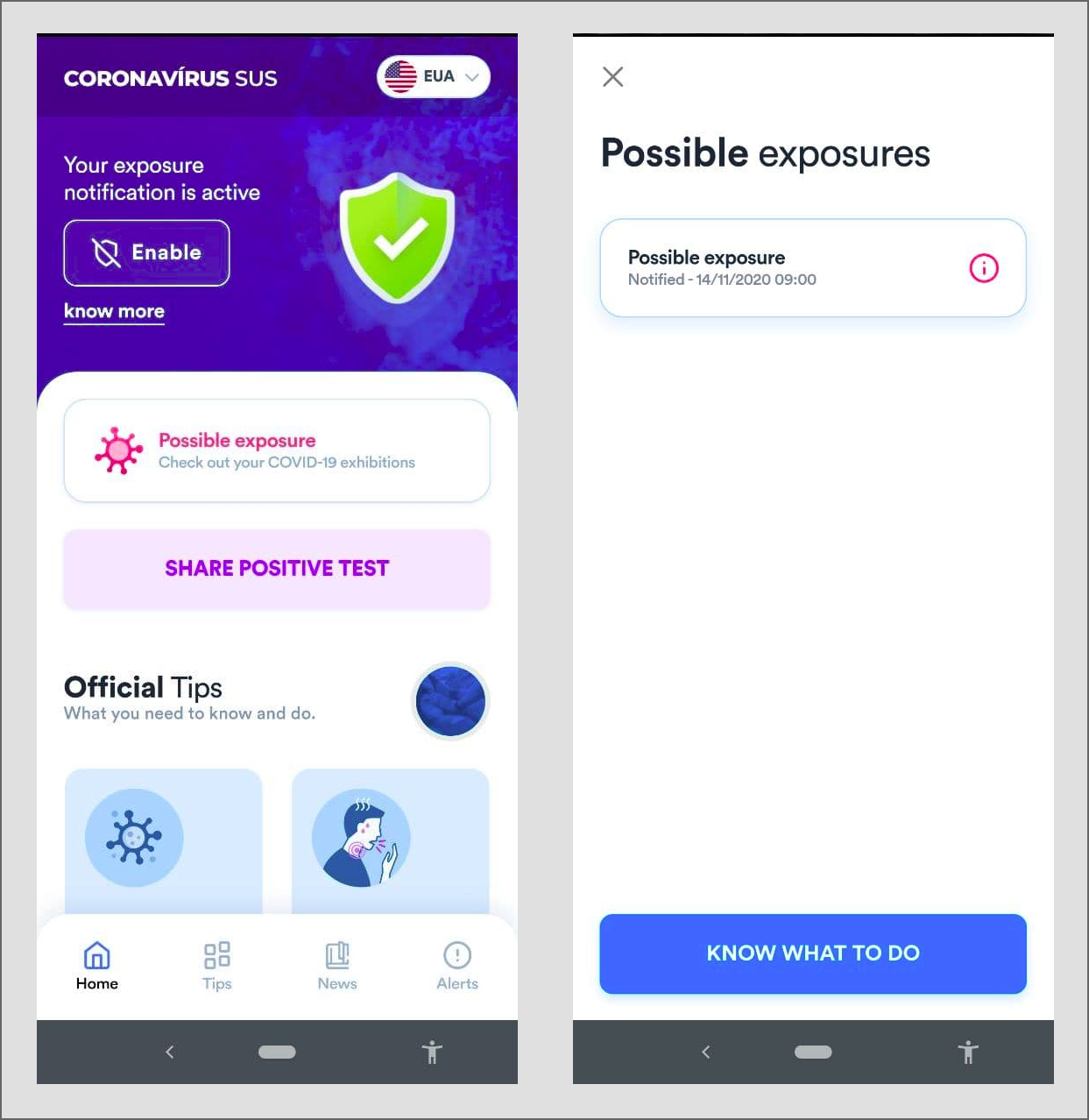}
  \caption{Example of contact notification. On the left, the app informs the contact-tracing feature is enabled and about potential exposure to COVID-19. There is also the option to share a positive test. On right, all the contact notifications the user had are displayed. The application is available in Portuguese, English, and Spanish.}
  \label{fig:app-sus}
\end{figure}

\textit{Coronav\'irus-SUS} can only work properly if the patient who tested positive for COVID agrees in sharing that result, like in any other contact tracing application. A positive patient must validate his/her test in a separate portal (http://validacovid.saude.gov.br) to prevent false positives and abuse. The system will cross-validate the patient's test with the data present in the public \textit{Rede Nacional de Dados em Sa\'ude} (RNDS, National Network of Health Data), where information about COVID-19 patients in Brazil is stored, and generate a token to be used for confirmation in the app. This portal has a separate privacy policy in compliance with Brazil's data protection regulation, LGPD\footnote{\url{https://validacovid.saude.gov.br/politica-privacidade}}. Image \ref{fig:app-sus} shows a contact notification as well as the "Validate Your Positive Test" option.

We present the information gathered from \textit{Fala.BR}, Brazil's platform for public information access. Any Brazilian citizen can request public information from the Federal Executive Office, according to the federal law no. 12.527 of November 18\textsuperscript{th}, 2011. The legal proceedings of our request for information, as well as the official answer, are publicly accessible (in portuguese) through \textit{Fala.BR} website\cite{falabr}. Table \ref{tab:data} resumes the obtained information. As one can see in this table, the adhesion of \textit{Coronav\'irus-SUS}, despite a favorable launching, is not yet expressive enough for a country with over 210 million people. Furthermore, active users have the option to not enable the contact tracing feature, so the reachability of the application is even smaller than the stated 1.81\%. Currently, the developers have no way to tell whether a user of \textit{Coronav\'irus-SUS} application has enabled the contact tracing feature in their smartphone because of privacy demands, as stated by DATASUS in our request for information available  \cite{falabr}. The extra proceedings for the validation of a positive result may be keeping the users from doing it, since only on November 10\textsuperscript{th}, 2020 Brazil had 23,973 new coronavirus cases, a number over 15 times greater than the total positive tests validated in the application.

\begin{table}
\centering
\begin{tabular}{ll} 
\hline
\textbf{Attribute}             & \textbf{Value}  \\ 
\hline
Total downloads                & 10.530.000      \\
Active users (Google Play)     & 1.520.000       \\
Active users (Apple Store)     & 2.290.00        \\
Penetration                    & 1.81\%          \\
Total of positive tests shared & 1.573           \\
\hline
\end{tabular}
\caption{Information about Coronav\'irus SUS adhesion, on the 10th of November, 2020.}
\label{tab:data}
\end{table}




Based on the previous analysis regarding the privacy of contact tracing applications, \textit{Coronav\'irus-SUS} functionality is less invasive and respects the privacy of its users regarding anonymity, decentralization, and storage of data. However, this decentralization impairs the developers' analytics to a certain degree, as stated by DATASUS \cite{falabr}.
The management issue might be keeping the \textit{Coronav\'irus-SUS} from reaching the intended population, since data analytics plays a considerable role in a country with considerable regional inequality of the size of Brazil. In addition, the lack of official advertisement of the \textit{Coronav\'irus-SUS} application from the Brazilian Government and from the major news outlets might be the main issue preventing its adhesion.
As the COVID-19 pandemic is still far from over, is still possible for \textit{Coronav\'irus-SUS} developers to achieve common ground between user privacy and application management, and for the Brazillian Government to run the necessary marketing campaign to ensure its reachability. 

\section{Key Research Challenges and Final Remarks}
\label{sec:conclusion}



The COVID-19 pandemic has caused directly millions of deaths all over the world as well as an important number of indirect ones, due to the exhaustion of the healthcare systems. Besides, the impact on the economy of many countries can decrease significantly the overall quality of life of people. In this context, Information and Communications Technologies (ICT) can be employed to control the impact of the pandemic. For example, contact tracing apps are expect to provide significant benefits in contagious disease. These apps can be viewed as a special kind of healthcare app that integrates specific communication technologies. In any case, more studies are needed to understand the impact of contact tracing on the evolution of the pandemic, since this will allow to include it in an evidence-based healthcare approach.


The rush for tools to face the spread of Sars-CoV-2 virus has fostered the rise of the several novel platforms for contact tracing. In this context, especially considering countries without public national healthcare systems, some regional entities (\textit{e.g.}, states and counties) are developing regional contact tracing apps, employing a decentralized approach. However, such decentralization hampers the execution of processing and analytics tasks. Even if the regional apps use the same APIs (such as the GAEN one), it is still unclear how the collected data could feed a national healthcare strategy. Thus, it is necessary to discuss how to employ federations of contact-tracing apps. A platform for these federations may present features, such as data deduplication, capacity optimization, and integrated analysis tools.


The data collected on contact tracing apps can be valuable for different healthcare related tasks. Thus, several different apps are expected to communicate with contact tracing ones. For example, remote patient monitoring and telemedicine apps could be enhanced with localization data. In any case, it would be necessary to decide how to implement privacy policies in order to protect the healthcare users. We envision that several organizations, besides the governmental ones, may be interested in the features supported by contact tracing. An example of these organizations is healthcare insurance. 


The research and development efforts to face the global COVID-19 health crisis will unveil new uses for contact tracing apps. Depending on the tasks and their intrinsic characteristics, such uses can impose challenges regarding privacy features. An example is the monitoring of Long COVID-19, also known as Chronic COVID Syndrome. In this case, the symptoms caused by the SARS-CoV-2 virus persist after the usual convalescence period. Thus, it is necessary to follow the use for a longer period. This long-term approach requires a longitudinal privacy analysis, since as more data is collected, more privacy concerns arrive.


As future work, we intend to enhance the evaluation of \textit{Coronav\'irus-SUS} as well as other relevant contact tracing apps. We plan to perform an in-depth forensic analysis of the Android version of \textit{Coronav\'irus-SUS} app. Furthermore, we are also looking at additional settings that could lead to important effects. For example, it is necessary to evaluate privacy features regarding the intersection of contact tracing and vaccination. In addition, it is necessary to propose guidelines for protecting users in order to prepare humanity to face similar challenges in future, without occurring in mass surveillance.

\bibliographystyle{IEEEtran}
\bibliography{biblio}


\begin{IEEEbiographynophoto}{Jéferson C. Nobre}
(jcnobre@inf.ufrgs.br) is Adjunct Professor at the Institute of Informatics, Federal University of Rio Grande do Sul (UFRGS). He holds a Ph.D. in Computer Science (UFRGS, 2015). Prof. Nobre has been involved in various research areas, mainly computer networks, network management, and computer system security. He has been publishing his works in important journals and conferences, such as the IEEE Communications Surveys and Tutorials, the IEEE Communications Magazine, the Journal of Network and Systems Management (Springer), the IFIP CNSM, IEEE/IFIP NOMS, and the IFIP/IEEE IM. He also served in the organization of scientific events and in the TPC of relevant symposia.
\end{IEEEbiographynophoto}

\begin{IEEEbiographynophoto}{Laura R. Soares}
(lrsoares@inf.ufrgs.br) is concluding the Bachelor degree in Computer Science at the Institute of Informatics, Federal University of Rio Grande do Sul (UFRGS). She has an Academic Research Scholarship from the National Council of Scientific and Technologic Development (CNPq). Her research areas of interest include computer networks, network management, and computer system security. She also volunteered at the organization of academic events in the University. 
\end{IEEEbiographynophoto}

\begin{IEEEbiographynophoto}{Briggette R. Huaytalla}
(borhuaytalla@inf.ufrgs.br) is a student in Master degree in Computer Science at the Institute of Informatics of the Federal University of Rio Grande do Sul (INF-UFRGS), Brazil, and member of the Computer Networks Group. Bachelor in Computer Science at National University of Engineering (UNI), Perú, and member of Scientific association specialized in computing (ACECOM-UNI). Briggette has an Academic Research Scholarship from the National Council of Scientific and Technologic Development (CNPq). Her research areas of interest include computer networks, computer system security, network security, application security, programmable networks and artificial intelligence.
\end{IEEEbiographynophoto}

\begin{IEEEbiographynophoto}{Elvandi S. Júnior} 
(elvandi.junior@inf.ufrgs.com) is Professor at the Vicente Dutra State Education Institute and Tutor at the University of Vale do Rio dos Sinos (UNISINOS). He holds a Ph.D. in Nanosciences (UFN, 2020), M.Sc. in Nanosciences (UFN, 2014), M.Sc. in Professional and Technological Education by the Federal University of Santa Maria (UFSM, 2019), Bachelor in Computer Science (UFN, 2012). Prof. Elvandi Júnior has been involved in several research areas, mainly computer networks, nanosciences, with an emphasis on microcontrollers and nanosensors, education and educational digital games.

\end{IEEEbiographynophoto}

\begin{IEEEbiographynophoto}{Lisandro Zambenedetti Granville} (granville@inf.ufrgs.br) is Full Professor at the Institute of Informatics of the Federal University of Rio Grande do Sul (INF-UFRGS), Brazil. Lisandro was chair of the IEEE ComSoc's Committee on Network Operations and Management (CNOM), co-chair of the IRTF's Network Management Research Group, and president of Brazilian Computer Society (SBC). He has served and general chair and TPC chair of conferences such as NOMS, IM, CNSM, and ICC, among others. Lisandro is editorial member of the Springer Journal of Network and Systems Management (JNSM) and IEEE Transactions on Network and Service Management (TNSM). He is also associate editor-in-chief of Wiley International Journal of Network Management (IJNM) and editor-in-chief of Springer Journal of Internet Services and Applications (JISA).

\end{IEEEbiographynophoto}

\end{document}